# BeCAPTCHA: Detecting Human Behavior in Smartphone Interaction using Multiple Inbuilt Sensors

**Alejandro Acien, Aythami Morales, Julian Fierrez, Ruben Vera-Rodriguez, Ivan Bartolome**

Biometrics and Data Pattern Analytics Lab (BiDA Lab), Autonomous University of Madrid, Madrid, Spain

{alejandro.acien, aythami.morales, julian.fierrez, ruben.vera, ivan.bartolome}@uam.com

**Abstract**

We introduce a novel multimodal mobile database called HuMIdb (Human Mobile Interaction database) that comprises 14 mobile sensors acquired from 600 users. The heterogeneous flow of data generated during the interaction with the smartphones can be used to model human behavior when interacting with the technology. Based on this new dataset, we explore the capacity of smartphone sensors to improve bot detection. We propose a CAPTCHA method based on the analysis of the information obtained during a single drag and drop task. We evaluate the method generating fake samples synthesized with Generative Adversarial Neural Networks and handcrafted methods. Our results suggest the potential of mobile sensors to characterize the human behavior and develop a new generation of CAPTCHAs.

## 1. Introduction

The research interest in smartphone devices has been constantly growing in the last years. The capacity of these devices to acquire, process, and storage a wide range of heterogeneous data offers many possibilities and research lines: user authentication (Frank et al. 2013; Patel et al. 2016; Voris et al. 2016; Monaco and Tappert 2018; Fierrez et al. 2018a), health monitoring (Albert et al. 2012; Arroyo-Gallego et al. 2017), behavior monitoring (Pei et al. 2010; Salem et al. 2014; Chen et al. 2015; Dua et al. 2019; Tavakolian et al. 2019), etc. Besides, the usage of mobile phones is ubiquitous. Mobile lines exceeded world population in 2018, and the amount of smartphones devices sold surpassed world population in 2014. This is one of the fastest growing manmade phenomenon ever, from 0 to 7.2 billion in barely three decades. In the same way, this widget has changed the way we access and create contents in internet. Recent surveys reveal that nearly three quarters (72.6%) of internet users will access the web via their smartphones by 2025. In fact, almost 51% of web access are actually made through mobile phones[1].

On the other hand, mobile web hazards are growing very fast as well. Malicious malware is also adapting to this new mobile era. Mobile bots employ the capacities of smartphones affecting multiples types of online services, such us: social media (e.g. mobile bots accounts propagate fake twitter messages (Chu et al. 2010), ticketing/travel, e-commerce, finance, gambling, ATO/Fraud, DDOs attacks, and price scrapping among others. According to (Distil 2018), these mobile bots use cellular networks by connecting through cellular gateways. Mobile bots can perform highly advanced attacks while remaining hidden in plain sight. In addition, they are very unlikely to be detected by IP address blocking. The Distil Research Lab showed that 5.8% of all mobile devices on cellular networks are used in malicious bot attacks (Distil 2018). In other study (Threat Metrix 2018), researchers reveal that mobile fraud reached 150 million global attacks in the first half of 2018 with attack rates rising 24% year-over-year.

In this context, new countermeasures against fraud adapted to mobile scenarios are necessary. One of the most popular methods to distinguish between humans and bots is known as CAPTCHA (Completely Automated Public Turing test to tell Computers and Humans Apart). These algorithms determine whether or not the user is human by presenting challenges associated to the cognitive capacities of the human beings. The most common challenges are: recognizing characters from a distorted image (text-based CAPTCHAs); identifying class-objects in a set of images (image-based CAPTCHAs); speech translation from distorted audios (sound-based audio CAPTCHAs); or newer systems that replace traditional cognitive tasks by a transparent algorithm capable of detecting bots and humans from their web behavior (reCAPTCHA 2019). However, recent advances in areas such as computer vision, speech recognition, or natural language processing have increased the vulnerabilities of CAPTCHA systems (Bursztein et al. 2011; Block et al. 2017; Akrout et al. 2019). Major advances in deep learning applied in those areas enable the generation of synthetic data of very natural appearance, therefore increasingly difficult to detect if used by bots.

---

[1] https://www.cnbc.com/2019/01/24/smartphones-72percent-of-people-will-use-only-mobile-for-internet-by-2025.html

| Ref. | Sensors | #Users | Sessions/user | Supervised | Public | Device | Task |
|---|---|---|---|---|---|---|---|
| Mahbub et al. 2016b | 13 (Key, Cam, Tou, Gyr, Acc, Mag, Lig, GPS, Blu, WiF, Pro, Tem, Pres) | 54 | ~248 | No | Yes | Shared | Free |
| Li and Bours 2018a | 2 (WiF, Acc) | 312 | 1 Month | No | No | Own | Free |
| Li and Bours 2018b | 2 (Acc, Gyr) | 304 | ~90 Sessions | No | No | Own | Free |
| Fridman et al. 2017 | 3 (App, Web, GPS) | 200 | 5 Months | No | No | Own | Free |
| Liu et al. 2018 | 5 (Tou, Pow, Acc, Gyr, Mag) | 10 | 3 Hours | Yes | No | Shared | Free |
| Shen et al. 2018 | 4 (Acc, Gyr, Mag, Ori) | 102 | 20 - 50 Days | No | Yes | Shared | Free |
| Deb et al. 2019 | 8 (Key, GPS, Acc, Gyr, Mag, Acc, Gra, Rot) | 37 | 15 Days | No | No | Own | Free |
| Tolosana et al. 2019 | 3 (Touch, Acc, Gyr) | 217 | ≤6 | No | Yes | Own | Fixed |
| **HuMIdb** | **14 (Acc, LAc, Gyr, Mag, Pro, Gra, Lig, Tou, GPS, WiF, Blu, Mic, Key, Ori)** | **600** | **≤5** | **No** | **Yes** | **Own** | **Fixed** |

**Table 1.** Summary of the existing multimodal mobile databases. Key- Keystroke, Acc- Accelerometer, Cam- Front camera, Touch- Touch gestures, Gyr- Gyroscope, Mag- Magnetometer, Prox- Proximity, Temp- Temperature, Press- Pressure, App- App usage, Web- Web browsing, Pow- Power consumption, Gra- Gravity, Rot- Rotation, LAc- Linear accelerometer, Mic- Microphone, Ori- Orientation. *Task* column shows whether the mobile sensors were recorded in the wild or while users completed prefixed tasks.

Most of the current CAPTCHAs have been designed to be used in a web interaction based on mouse and keyboard interfaces. In this paper we explore the potential of mobile devices to detect human-machine interaction. The main contributions are twofold: i) a new public HuMIdb (Human Mobile Interaction database) that characterize the interaction of 600 users according to 14 sensors during normal human-mobile interactions in an unsupervised scenario with more than 200 different devices; ii) we propose a new Behavioral CAPTCHA system (called BeCAPTCHA[2]) based on modelling the user behavior in smartphone interaction using multiple inbuilt sensors.

### 1.1. Multimodal Mobile Datasets

Table 1 summarizes previous multimodal mobile databases. UMDAA-02 (Mahbub et al. 2016b) is a multimodal mobile database that includes 14 mobile sensors: front camera, touchscreen, gyroscope, accelerometer, magnetometer, light sensor, GPS, Bluetooth, WiFi, proximity sensor, temperature sensor, and pressure sensor. The data was collected during 2 months from 48 volunteers in an unsupervised scenario with 248 sessions per user in average and using the same smartphone (Nexus 5). In (Li and Bours 2018a), WiFi and accelerometer mobile signals were collected from 312 participants in an unsupervised scenario during a period of 1 month. In (Li and Bours 2018b), the same authors captured accelerometer and gyroscope mobile data from 304 participants in an unsupervised scenario with 90 sessions per user in average. Fridman et al. (2017) collected a multimodal mobile dataset that contains stylometry, app usage, web browsing, and GPS behavioral biometrics data from 200 subjects over a period of 5 months. They launched a mobile application that acquired the mobile data during natural mobile-human interactions (unsupervised scenario). They also mention that the main problem in the acquisition process was the battery drain, as the app was recording all the time the smartphone was on. In other work (Liu et al. 2018), the authors collect touch gestures, power consumption, accelerometer, gyroscope, and magnetometer mobile signals from 10 participants under laboratory conditions and with the same mobile device (supervised scenario) during a period of three hours. Shen et al. (2018) collected accelerometer, gyroscope, orientation, and magnetometer mobile sensors from 102 subjects for active mobile authentication in an unsupervised scenario. During the acquisition they covered three smartphone-operating environments: hand-hold, table-hold, and hand-hold-walking. In a recent work (Deb et al. 2019), the authors collected up to 30 mobile sensor signals from 37 volunteers over a period of 15 days with an Android application that works in background and acquire the mobile data passively. Finally, Tolosana et al. (2019) collect a database of mobile touch on-line data named MobileTouchDB. The database is focused in mobile touch patterns and contains more than 64K on-line character samples performed by 217 users with a total of 6 sessions. They also acquired accelerometer and gyroscope signals under unsupervised conditions.

---
[2] Spanish Patent Application P202030066

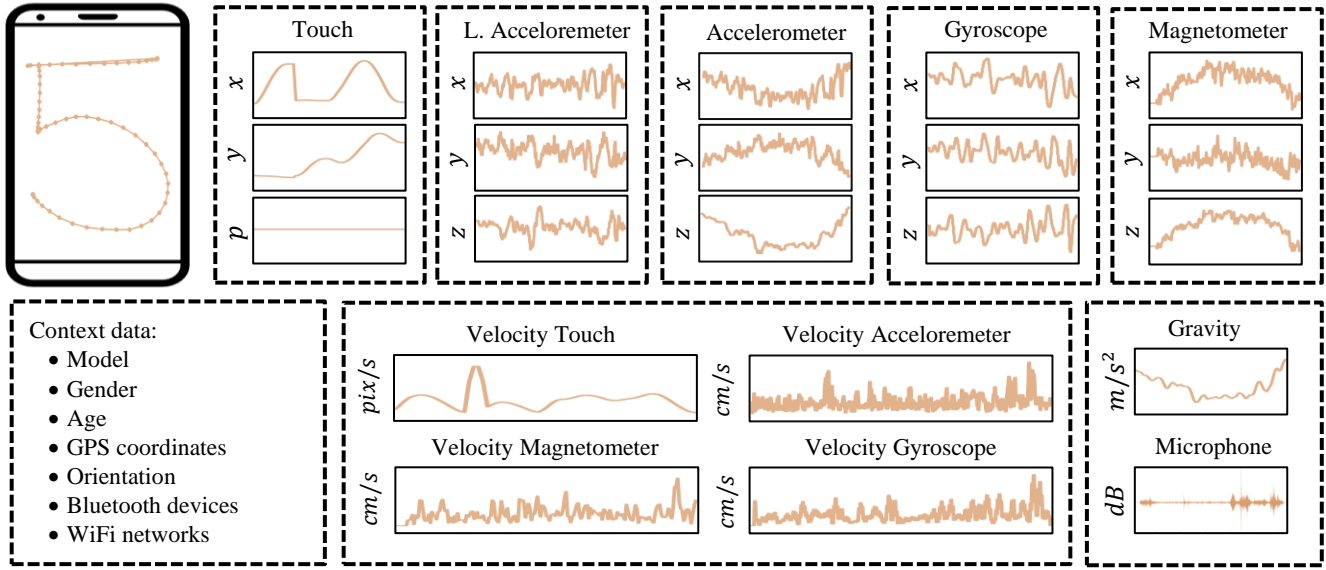

**Figure 1.** Full set of data generated during one of the HuMIdb task.

## 2. The HuMI Database

In this section we introduce the Human Mobile Interaction database (HuMIdb), a novel multimodal mobile database that comprises more than 5 GB from a wide range of mobile sensors acquired under unsupervised scenario. The database includes 14 sensors (see Table 2 for the details) during natural human-mobile interaction performed by 600 users. For the acquisition, we implemented an Android application that collects the sensor signals while the users complete 8 simple tasks with their own smartphones and without any supervision whatsoever (i.e., the users could be standing, sitting, walking, indoors, outdoors, at daytime or night, etc.). The different tasks are designed to reflect the most common interaction with mobile devices: keystroke (name, surname, and a pre-defined sentence), tap (press a sequence of buttons), swipe (up and down directions), air movements (circle and cross gestures in the air), handwriting (digits from 0 to 9), and voice (record the sentence *I'm not a robot*). Additionally, there is a drag and drop button between tasks.

The acquisition protocol comprises 5 sessions with at least 1 day gap among them. It is important to highlight that in all sessions, the 1-day gap refers to the minimum time between one user finishes a session and the next time the app allows to have the next session. At the beginning of each task, the app shows a brief pop up message explaining the procedure to complete each task. The application also captures the orientation (landscape/portrait) of the smartphone, the screen size, resolution, the model of the device, and the date when the session was captured.

Regarding the age distribution, 25.6% of the users were younger than 20 years old, 49.4% are between 20 and 30 years old, 19.2% between 30 and 50 years old, and the remaining 5.8% are older than 50 years old. Regarding the gender, 66.5% of the participants were males, 32.8% females, and 0.7% others. Participants performed the tasks from 14 different countries (52.2%/47.0%/0.8% from Europe, America, and Asia respectively) using 278 different devices. The database was captured using an app developed specifically for this work. All users were informed about the sensors and information captured, and all users have signed the necessary consent agreement. The information is captured only during the usage of the app and all data were anonymized.

Figure 1 shows an example of the handwriting task (for digit "5") and the information collected during the task. Note how a simple task can generate a heterogeneous flow of information related with the user behavior: the way the user holds the device, the power and velocity of the gesture, the place, etc.

| Sensors | Sampling Rate | Features | Power Consumption |
|---|---|---|---|
| Accelerometer | 200 Hz | $x, y, z$ | Low |
| L.Accelerometer | 200 Hz | $x, y, z$ | Low |
| Gyroscope | 200 Hz | $x, y, z$ | Low |
| Magnetometer | 200 Hz | $x, y, z$ | Low |
| Orientation | NA | $l$ or $p$ | Low |
| Proximity | NA | $cm$ | Low |
| Gravity | NA | $m/s^2$ | Low |
| Light | NA | $lux$ | Low |
| TouchScreen | E | $x, y, p$ | Medium |
| Keystroke | E | $key, p$ | Medium |
| GPS | NA | Lat., Lon., Alt., Bearing, Accuracy | Medium |
| WiFi | NA | SSID, Level, Info, Channel, Frequency | High |
| Bluetooth | NA | SSID, MAC | Medium |
| Microphone | 8 KHz | Audio | High |

**Table 2.** Description of all sensor signals captured in HuMIdb. E=Event-based acquisition. The timestamp parameter is captured for all sensors.

## 3. BeCAPTCHA: Detecting Human Behavior from Mobile Inbuilt Sensors

HuMIdb offers the opportunity to model human behavior. Among the multiple applications, in this work we will explore the use of human interaction to develop a new generation of CAPTCHA systems based on mobile inbuilt sensors. We propose to employ the right swipe gesture captured at the end of each HuMIdb task (drag and drop action when the user scrolls the *Next* button to the right). We propose to model the gesture according to features obtained from the touchscreen and accelerometer. To evaluate the CAPTCHA, we will employ human samples (from HuMIdb) and synthetic ones (bot-like samples). We assume a challenging scenario where the attacker (malicious bot developer) can generate synthetic gestures trying to mimic the sensor signals generated during human-mobile interaction.

| Parameters | Description |
|---|---|
| Duration ($D$) | $t_N - t_0$ |
| Distance ($L$) | $\|(x_N, y_N) - (x_0, y_0)\|$ |
| Displacement ($P$) | $\sum_{i=0}^{N-1} \|(x_{i+1} - x_i) - (y_{i+1} - y_i)\|$ |
| Angle ($\alpha$) | $\tan^{-1}(|(x_N - x_0)|/|(y_N - y_0)|)$ |
| Mean velocity ($V$) | $\frac{1}{N}\sum_{i=0}^{N-1} \|(x_{i+1} - x_i) - (y_{i+1} - y_i)\|/(t_{i+1} - t_i)$ |
| Move Efficiency ($E$) | $P/L$ |

**Table 3.** Touch features extracted for the characterization of the gestures.

### 3.1. Feature Extraction: Characterizing Swipe Gestures

To characterize swipe gestures from the touchscreen and accelerometer signals, we have adapted two feature sets previously employed in (Li and Bours 2018; Chu et al. 2018) for bot detection and user authentication respectively.

The interaction of the user with the Touch screen is defined by a time sequence $\mathbf{s}_T = \{\mathbf{x}, \mathbf{y}, \mathbf{p}, \mathbf{t}\}$ with length $N$, composed by the coordinates $\{\mathbf{x}, \mathbf{y}\}$, the pressure $\mathbf{p}$ (when available), and the timestamp $\mathbf{t}$. First, the coordinates $\{\mathbf{x}, \mathbf{y}\}$ are normalized by the size of the screen. Second, the pressure is discarded as it is not available in most of the devices. Third, the touch features consist of six global features defined in Table 3.

The accelerometer signal is defined by a sequence $\mathbf{s}_A = \{\mathbf{x}, \mathbf{y}, \mathbf{z}, \mathbf{t}\}$. The feature set chosen for the accelerometer signal was adapted from (Li and Bours 2018), in which they calculate the mean, median, root-mean-square, and standard deviation of the three accelerometer axes $\{\mathbf{x}, \mathbf{y}, \mathbf{z}\}$ for user authentication.

### 3.2. Generating Human-Like Gestures: Fakes

A swipe gestures can be defined by a spatial trajectory (sequence of points $\{\mathbf{x}, \mathbf{y}\}$) and a velocity profile determined by the timestamp sequence $\mathbf{t}$. To generate synthetic swipe patterns, we will follow two approaches: handcrafted synthesis and Generative Adversarial Network (GAN) synthesis.

*Method 1: Handcrafted Synthesis*
We observed that most of the human swipe trajectories obtained from our drag and drop task are linear. The handcrafted approach generates swipe trajectories according to a straight line shape and a realistic velocity profile. For this, we first estimate the probability distribution of length and angle of human swipe gestures in HuMIdb. Note that the size and coordinates of each human swipe varies depending on the device features so we have normalized each one by the total size of the screen.

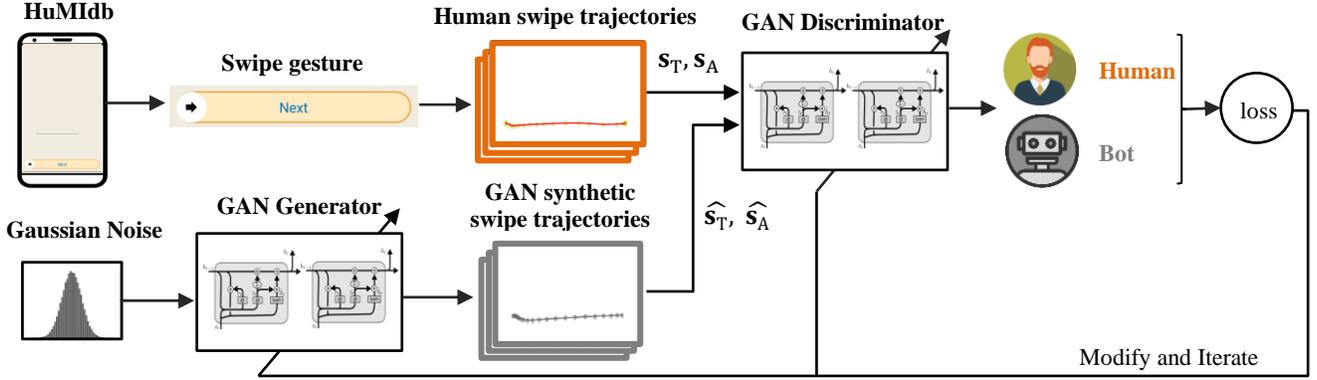

**Figure 2.** The proposed architecture to train a GAN Generator of synthetic swipe gestures characterized by touch $\widehat{s_T}$ and accelerometer $\widehat{s_A}$ sequences. The Generator learns the human features of the swipe gestures and generate human-like ones from Gaussian Noise and human sequences $s_T$, $s_A$.

The synthetic trajectories are defined by the initial point $(x_0, y_0)$, duration $(t_N - t_0)$, angle ($\alpha$), and the velocity profile $\{v, t\}$. We have synthesized the fake trajectories according to distributions of these parameters fitted from human data (except for the velocity profile). With the aim to emulate human behaviors, we spaced the points of the linear trajectory on a log scale (emulating a velocity profile with the initial acceleration observed in human samples).

The accelerometer signals are synthesized as random sequences generated from a Gaussian distribution with mean and standard deviation estimated from real accelerometer signals from HuMIdb.

*Method 2: GAN Synthesis*
For this approach, we employ a GAN (Generative Adversarial Network) architecture firstly proposed by Goodfellow et al. (2014), in which two neuronal networks, commonly named Generator and Discriminator, are trained in adversarial mode. The Generator tries to fool the Discriminator by generating fake samples (touch trajectories and accelerometer signals in this work) very similar to the real ones, while the Discriminator has to discriminate between the real samples and the fake ones created (see Figure 2 for the details). Once the Generator is trained, then we can use it to synthesize swipe trajectories very similar to the real ones.

The topology employed in both Generator and Discriminator consist of two LSTM (Long Short-Term Memory) layers followed by a dense layer, very similar to a recurrent auto-encoder. The LSTM layers learn the time relationships of human swipe sequences, while the dense layer is used as a classification layer to distinguish between fake and real swipe trajectories in the Discriminator or to build synthetic mouse trajectories in the Generator. To synthesize accelerometer signals, we follow the same GAN architecture described before, but extending the input of the generator from $\{x, y\}$ swipe coordinates to $\{x, y, z\}$ accelerometer axes.

### 3.3. Experimental Protocol
Both GAN networks were trained using more than 10K human samples extracted from the HuMIdb. Training details: learning rate $\alpha = 2 \cdot 10^{-4}$, Adam optimizer with $\beta_1 = 0.5$, $\beta_2 = 0.999$, and $\varepsilon = 10^{-8}$. The system was trained for 50 epochs with a batch size of 128 samples for both Generator and Discriminator. The loss function was '*binary crossentropy*' for the Discriminator and '*mean square error*' for the Generator. The model was trained and tested in *Keras-Tensorflow*.

We generated 12K synthetic samples according to the two methods proposed. Once we have extracted the global features from human and synthetic swipe trajectories (up to 30K trajectories between both groups), we employ a SVM (Support Vector Machine) classifier with a RBF (Radial Basis Function). The experiments are divided into two different scenarios depending on the training data employed: once-class or multiclass. In the one-class scenario, we train the SVM classifier using only the human samples and test with both human samples and synthetic ones, in order to study whether the classifier is able to detect bots as abnormal human behavior. In the multiclass scenario, we train and test the SVM classifier with both human and synthetic samples in order to analyze whether the classifier is able to find discriminative features between both groups.

In both scenarios, there is not overlap between train and test sets. For each SVM, we train the classifier by using the 70% of all samples and test with the 30% remaining (randomly chosen). Each experiment was repeated 5 times and the results were computed as the average of the 5 iterations.

### 4. Results and Discussion

Table 4 shows the bot detection errors for the different synthetic trajectories (in columns) generated in this work when comparing with the human ones. The results are presented in terms of EER (Equal Error Rate) defined as the point

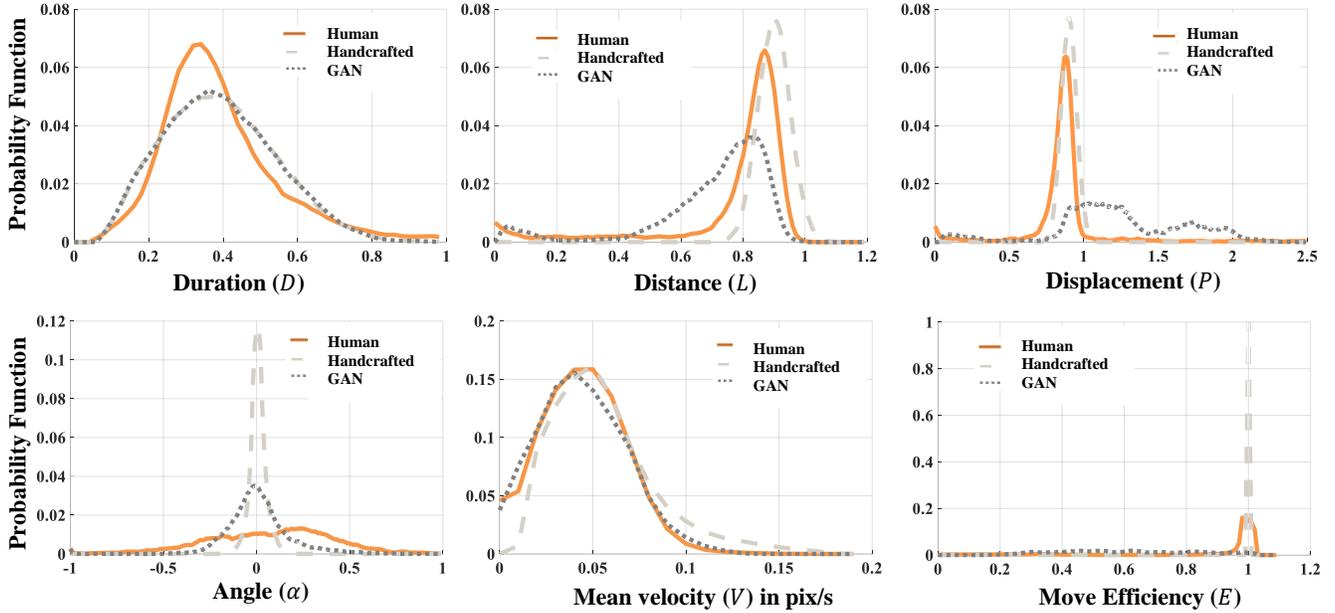

**Figure 3.** Probability functions of the six global features for human, handcrafted, and GAN swipe trajectories.

where FMR (False Match Rate) and FNMR (False Non-Match Rate) are equal.

We can observe that the results achieved for the one-class classification scenario are significantly worse (higher errors) than those achieved in multiclass classification, as expected because synthetic samples are not considered when training the classifier. For one-class classification, the synthetic samples generated with the GAN network are capable of fooling the classifier more times compared to the other types of synthetic samples. The same applies to multiclass training.

To better understand the results, Figure 3 shows the probability functions of the six features proposed for the three types of touch signals (i.e. humans and both synthetic generation methods). Synthetic distributions do not completely fit the human distributions, but they present a behavior similar to the human samples. First, we can observe that the Move Efficiency of the handcrafted trajectories is equal to 1, this happens because in swipe trajectories with straight line shape the distance and displacement are equal. This is the reason why the one-class classifier detects these synthetic trajectories so easily. Note that the Duration (length) of both handcrafted and GAN synthetic swipes were computed as a Gaussian distribution with the same mean and standard deviation as the human ones so both probability distributions are equal. Regarding Distance and Displacement, the GAN trajectories fit worse than the handcrafted ones. We suggest that the main reason for this is that the GAN network generates smoother swipe trajectories than the human ones without abrupt direction changes, causing longer displacements in less distance (like a parabolic function). Finally, the Velocity Profile of both synthetic swipe trajectories are very similar to the human ones, the initial acceleration applied to the function-based trajectories reproduces human behaviors with great similarity while the GAN network learns very realistic Velocity Profiles of human swipe trajectories as well.

|  | Bot Generation | |
|---|---|---|
| **SVM Training** | **Handcrafted** | **GAN** |
| One-class (Tou) | 38.3 | 46.7 |
| Multiclass (Tou) | 2.2 | 4.1 |
| One-class (Tou+Acc) | 0 | 23.2 |
| Multiclass (Tou+Acc) | 0 | 0.8 |
| Cross-Multiclass (Tou+Acc) | 10.1 | 7.2 |

**Table 4.** Equal Error Rate (%) in bot detection for the different scenarios including human and synthetic samples (i.e. Handcrafted, and GAN). In Cross-Multiclass experiments, the SVM is trained with fake samples of one generation method, and evaluated with the second one (e.g. the 10.1% was obtained training with HC samples and testing with GAN samples).

The combination of touch and accelerometer features outperform the previous results with more than 50% error reduction in bot detection. These results suggest the potential of multimodal approaches. Finally, Table 4 shows the performance when the generation method of fake samples used for training and testing are different. As it is expected, the EER increases when training samples does not include the same generation method employed for the evaluation. However, the system is capable of maintaining a correct classification rate around 90%.

## 5. Conclusions

In this paper we have explored the potential of mobile devices to model human-machine interaction. We present a novel multimodal mobile database HuMIdb that comprises 14 mobile sensors captured from 600 users in an unsupervised scenario. Although in this paper we focus on the HuMIdb for bot detection, this new dataset offers multiple research opportunities and challenges.

We introduce a new CAPTCHA system based on the analysis of signals from inbuilt sensors. This behavioral information acquired through mobile sensors describe inner human features, such as neuromotor abilities, cognitive skills, human routines, and habits. All these patterns can help to develop new bot detection algorithms for mobile scenarios. We have evaluated the method using both human swipes trajectories (HuMIdb samples) and synthetic (bot samples). The results obtained reveal the potential of these signals to model human device interaction. Future works will include new approaches that exploit complementarity between tasks and sensors, and smart fusion to exploit heterogeneity of the data (Fierrez et al. 2018b).

## Acknowledgments

This work has been supported by projects: BIBECA (RTI2018-101248-B-I00 MINECO/FEDER), and BioGuard (Ayudas Fundación BBVA a Equipos de Investigación Científica 2017). Spanish Patent Application P202030066.